\documentclass[12pt]{article}
\usepackage{graphicx, here, amsfonts, amssymb, epsfig, hhline}
\def\Journal#1#2#3#4{{#1} {\bf #2} (#4) #3}

\def\NPB{{\em Nucl. Phys.} B}
\def\NPBP{{\em Nucl. Phys. (Proc. Suppl.)} B}
\def\PLB{{\em Phys. Lett.}  B}
\def\PRL{\em Phys. Rev. Lett.}
\def\PRD{{\em Phys. Rev.} D}
\def\ZPC{{\em Z. Phys.} C}
\def\EPG{{\em Eur.Phys.J.} C}
\def\EPJL{{\em Europhys. Lett.}}
\def\IJMP{\em Int. Jour. of Mod. Phys.}

\def\be{\begin{equation}}
\def\ee{\end{equation}}
\def\bea{\begin{eqnarray}}
\def\eea{\end{eqnarray}}

\begin{document}
\begin{center}
{\Large \bf $p p$ Elastic Scattering at LHC and Nucleon Structure   }

\vskip 0.5cm 
 M. M. Islam$^{a,}$\footnote{\it E-mail:
islam@phys.uconn.edu}, R. J. Luddy$^{a,}$\footnote{\it E-mail:
rjluddy@attbi.com} 
and
A. V. Prokudin$^{b,c,}$\footnote{\it E-mail: prokudin@to.infn.it}
\vskip 0.5cm
{\small\it
(a) Department of Physics,  University of Connecticut, 
Storrs, CT 06269, USA 
}
{\small\it
\vskip 0.cm
(b)  Dipartimento di Fisica Teorica,
Universit\`a Degli Studi Di Torino, 
Via Pietro Giuria 1,
10125 Torino, ITALY
and
Sezione INFN di Torino, ITALY}
\vskip 0.cm
{\small\it
(c) Institute For High Energy Physics,
142281 Protvino,  RUSSIA}

\vskip -0.2cm
\parbox[t]{14.cm}{\footnotesize
\begin{center}
{\bf Abstract}
\end{center}                                              
     High energy elastic  $p p$ scattering at the Large Hadron Collider (LHC) at 
c.m. energy 14 TeV is predicted using the asymptotic behavior of $\sigma_{tot}(s)$
and $\rho(s)$ known 
from dispersion relation calculations and the measured elastic $\bar p p$ 
differential cross section at $\sqrt{s} = 546 {\rm GeV}$. 
The effective field theory model underlying the phenomenological analysis 
describes the nucleon as having an outer cloud of quark-antiquark condensed 
ground state, an inner core of topological baryonic charge of radius 
$\simeq 0.44F$ and a still smaller valence quark-bag of radius $\lesssim 0.1\;{\rm F}$. The 
LHC experiment 
TOTEM (Total and Elastic Measurement), if carried out with sufficient precision 
from $|t| = 0$ to $|t| > 10\; {\rm GeV^2}$, will be able to test this structure 
of the nucleon.}
\end{center}

\vskip 0.7cm

     High-energy elastic $\bar p p$ and  $p p$ scattering have been 
measured at CERN ISR \cite{[1]} and SPS Collider \cite{[2]} and at Fermilab \cite{[3]},
\cite{[3a]}  over 
the energy range $\sqrt{s}= 23.5\;{\rm GeV}$ to $1.8 \;{\rm TeV}$. These measurements provide 
a broad perspective of the energy dependence of elastic differential cross sections 
and of the asymptotic behavior of total cross section $\sigma_{tot}(s)$ and of the 
ratio of real to imaginary part of the forward amplitude $\rho(s)$. They have led to various 
phenomenological models of elastic scattering, such as (1) impact-picture model \cite{[4]},
(2) Regge pole-cut models \cite{[5]},\cite{[threepomerons]}, (3) QCD motivated eikonal model \cite{[6]}, 
(4) cloud-core model \cite{[7]}. Theoretical investigation of the last model has shown 
that  -- an effective field theory model with quarks interacting via a scalar field underlies 
it \cite{[8]}. Soft diffractive processes mediated by 
Pomerons including high-mass diffractive dissciation 
have also been proposed to describe elastic 
scattering at LHC for small $|t|\lesssim 0.5\;{\rm GeV^2}$ \cite{[8a]}.

Study of the field theory model further indicates that the nucleon has an outer cloud of quark-antiquark 
condensed ground state analogous to a superconducting state, an inner core of topological baryonic 
charge and a still smaller quark-bag of valence quarks. The elastic scattering experiment TOTEM at LHC 
\cite{[9]}, if carried out with sufficient precision, will be able to identify these three regions 
inside the nucleon by their characteristic behaviors reflected in the elastic differential cross section. 
The present investigation reports our quantitative prediction of ${d\sigma}/{dt}$ at LHC at $\sqrt{s}=14\;{\rm TeV}$. 
Preliminary calculation of this differential cross section was reported earlier \cite{[10]}. We show 
how the behavior of ${d\sigma}/{dt}$ in different ranges of $|t|$ relates with the three 
regions inside the nucleon. 
The parameters in our model are determined such that the high-energy asymptotic behavior of $\sigma_{tot}(s)$ and $\rho(s)$ 
based on dispersion relation \cite{[11]} and the measured elastic $\bar p p$  differential cross section at 
$\sqrt{s}=546\;{\rm GeV}$ \cite{[2]} are satisfactorily described. General asymptotic requirements of 
the diffraction amplitude are shown to be satisfied by our phenomenological diffraction amplitude. 
We also present our predicted $p p$  elastic differential cross section at $\sqrt{s}=500\;{\rm GeV}$
 -- measurement of which is planned at the Relativistic Heavy Ion Collider \cite{[12]}.
     
We describe the crossing even ($C$-even) diffraction amplitude using the 
impact parameter representation 
\be
T^+_D(s,t)=ipW\int^{\infty}_{0}bdb J_0(bq)\Gamma^+_D(s,b)
\label{1}
\ee
\noindent                                                            
with a profile function \cite{[7]}
\be
\displaystyle \Gamma^+_D(s,b)=g(s)\Big[\frac{1}{1+exp[(b-R)/a]}+\frac{1}{1+exp[-(b+R)/a]}-1\Big]\; ,
\label{2}
\ee                           
where $W=\sqrt{s}$, $q=\sqrt{|t|}$, $R$ and $a$ are energy dependent:$R\equiv R(s)=R_0+R_1(\ln s - i\pi/2)$,
$a\equiv a(s)=a_0+a_1(\ln s - i\pi/2)$.
$g(s)$ is a crossing even function: $g^*(se^{i\pi}) = g(s)$ which asymptotically
 goes to a real positive constant. Besides the diffraction amplitude, the model has a 
hard scattering amplitude originating from one nucleon core scattering off the other core via 
vector meson $\omega$ exchange, while their outer clouds overlap and interact independently. 
The hard scattering amplitude is given by an amplitude $T_1(s,t)$ due to a single hard 
collision multiplied by an absorption factor $[1-\Gamma^{\bar pp}_D(s,0)]$ or $[1-\Gamma^{pp}_D(s,0)]$. 
In the absorption factor, the possibility of a $C$-odd contribution at zero impact parameter is taken 
into account, so that $\Gamma^{\bar pp}_D(s,0)=\Gamma^+_D(s,0)+\Gamma^-_D(s,0)$, 
$\Gamma^{pp}_D(s,0)=\Gamma^+_D(s,0)-\Gamma^-_D(s,0)$. 

     The diffraction amplitude obtained from Eqs.(\ref{1}) and (\ref{2}) satisfies the general properties associated with the phenomenon of diffraction:
\begin{enumerate}
\item It leads to $\sigma_{tot}(s)\sim (a_0+a_1 \ln s)^2$, i.e. qualitative saturation of 
Froissart-Martin bound.
\item It yields  $\rho(s)\simeq {\pi a_1}/{(a_0+a_1 \ln s)}$ asymptotically, so that the derivative 
dispersion relation result \cite{[13]}  $\rho(s)= \pi/\ln s $ is satisfied.
\item It obeys the Auberson-Kinoshita-Martin scaling, i.e. $T^+_D(s,t)\sim is \ln^2s f(|t|\ln^2s)$
asymptotically.
\item It is $C$-even, and therefore yields equal $\bar p p$ and  $p p$
  total and differential cross sections.
\end{enumerate} 
\noindent
The asymptotic properties 1-3 of the diffraction amplitude $T^+_D(s,t)$ can be seen 
in the following way:  We express the profile function $\Gamma^+_D(s,b)$ in the form 
\be
\Gamma^+_D(s,b) = g(s)\frac{\sinh R/a}{\cosh R/a+\cosh b/a}\; .
\label{3}
\ee
\noindent                                                             
Then we change the variable of integration from $b$ to $\zeta=b/a$ and rotate the line 
of integration a little to the real axis. This leads to
\be
\displaystyle T^+_D(s,t) = i p W g(s) a^2\int^{\infty}_{0}
\zeta d\zeta J_0(\zeta q a)\frac{\sinh R/a}{\cosh R/a+\cosh \zeta}\; .
\label{4}
\ee   
We observe that, when $a_0+a_1 \ln s \rightarrow \infty$,
\be
\frac{R_0+R_1(\ln s -i\pi/2)}{a_0+a_1(\ln s -i\pi/2)}=
\frac{R_0-ra_0}{a_0+a_1(\ln s -i\pi/2)}+r\simeq r,
\label{5}
\ee
where  $r\equiv R_1/a_1$ is a real quantity. Using this in the integrand 
in (\ref{4}), we obtain the first three properties. The fourth one follows 
from the $C$-even form (\ref{2}) of the profile function. We note that the 
$\ln s$ dependence in the diffraction amplitude always 
occurs in the combination $(a_0+a_1\ln s)$, which is independent of the scale of $s$.

For $|t|\ne 0$, $T^+_D(s,t)$ is given by the crossing symmetric form \cite{[7]}
\be
\displaystyle T^+_D(s,t)\simeq is g(s)a\frac{1}{2}\{-\pi i (R+i\pi a)H^{(1)}_0[q(R+i\pi a)]+\pi i(R-i\pi a)
H^{(2)}_0[q(R-i\pi a)]\}\; .
\label{6}
\ee
When $ q|R\pm i\pi a|\gg 1$, the Hankel functions fall off exponentially,
 which leads to ${|T^+_D(s,t)|}/{s}\sim exp[-q\pi (a_0+a_1\ln s)]$ as $s\rightarrow \infty$. 
Since $d\sigma /dt = 4 \pi |T(s,t)/s|^2$, we find that for $q$ fixed and $s\rightarrow \infty$, the differential cross section due to 
diffraction vanishes. 
On the other hand, the total cross section due to 
diffraction tends to infinity as $\ln^2 s$. Hence, Martin's theorem \cite{[14]} predicts a 
zero of ${\rm Re}T^+_D(s,t)$ in the near forward direction. Indeed, in our calculations we find
 such zeros indicating further the asymptotic nature of our diffraction amplitude.

     To extend our previous calculations to the LHC energy, we needed a way to determine the $s$
 dependence of $g(s)$. Accurate model independent analysis by Kundrat and Lokajicek \cite{[15]} 
has shown that inelasticity due to diffraction at $b = 0$, i.e. $|exp[i\chi^+_D(s,0)]|$ is small 
but finite at high energy and decreases slowly with increasing $s$. This has led us to consider 
the following simple crossing even parameterization:
\be
e^{i\chi^+_D(s,0)}=\eta_0+\frac{c_0}{(se^{-i\pi/2})^\sigma}\; .
\label{7}
\ee
As $g(s)$ is related to $exp[i\chi^+_D(s,0)]=1-\Gamma^+_D(s,0)$  via Eq.(\ref{3}), the three 
energy independent parameters in (\ref{7}) together with other diffraction parameters 
allow us to obtain $g(s)$ at different values of $s$. To take into account the energy 
dependence of $\Gamma^-_D(s,0)$, we have assumed a similar
 parameterization:
\be
\Gamma^-_D(s,0)=i\lambda_0-i\frac{d_0}{(se^{-i\pi/2})^\alpha}\; ,
\label{8}
\ee
which of course has the required crossing odd property. Finally, for the energy 
dependence of the term $\hat \gamma (s) exp[i\hat\theta(s)]= \tilde \gamma exp[i\hat\chi(s,0)]$, 
we have taken the parameterization:
\be
\hat \gamma (s) exp[i\hat\theta(s)]=\hat\gamma_0+\frac{\hat\gamma_1}{(se^{-i\pi/2})^{\hat\sigma}}\; .
\label{9}
\ee
We note an odderon contribution (an asymptotic crossing-odd amplitude) in our analysis:
\be
\pm\eta_0s\hat\gamma_0\frac{F^2(t)}{m^2-t} \;(\rm +\; sign\; for \;\bar pp, -\;  sign\; for\;pp),
\label{10}
\ee
where $F(t)$ is the $\omega NN$ form factor. As in the previous work \cite{[7]}, we take
\be
F^2(t) = \beta(m^2-t)^{1/2}K_1[\beta(m^2-t)^{1/2}]\; .
\label{10a}
\ee
This corresponds to a smoothed Yukawa potential  
$e^{-m(r^2+\beta^2)^{1/2}}/(r^2+\beta^2)^{1/2}$ between nucleons due to 
$\omega$ exchange and leads to an exponential fall-off 
( Orear fall-off ) of the differential cross section:
$d\sigma /dt \sim exp(-2\beta\sqrt{|t|})$, 
when hard scattering dominates.

     There are now thirteen energy-independent parameters in our model: $R_0$, $R_1$, $a_0$, $a_1$, 
$\eta_0$, $c_0$, $\sigma$, $\lambda_0$, $d_0$, $\alpha$, $\hat\gamma_0$, $\hat\gamma_1$, $\hat\sigma$. 
 As mentioned earlier, we determine these parameters 
by requiring that they satisfactorily describe the high energy asymptotic behavior of $\sigma_{tot}(s)$ and
$\rho(s)$  given by dispersion relation calculations \cite{[11]} and the experimental
$\bar pp$  elastic differential cross section at $\sqrt{s}=546\;{\rm GeV}$ \cite{[2]}. 
The values of the parameters obtained by us are:
 $R_0= 2.50$, $R_1= 0.0385$, $a_0= 0.57$, $a_1=0.106$, 
$\eta_0= 0.02$, $c_0= 1.40$, $\sigma= 0.50$, $\lambda_0= 0.375$, $d_0= 13.0$, $\alpha= 0.38$, 
$\hat\gamma_0= 2.81$, 
$\hat\gamma_1=400.0$, $\hat\sigma=1.30$. (Our unit of energy
is $1$ GeV.) $\beta$ and $m$ are not free parameters in our present analysis. They are kept fixed at values found previously: $\beta= 3.075$, $m = 0.801$, which correspond to a radius $0.44 \; \rm F$ for the nucleon repulsive core.

\begin{figure}[H]
\centering {\epsfxsize=170mm
\epsffile{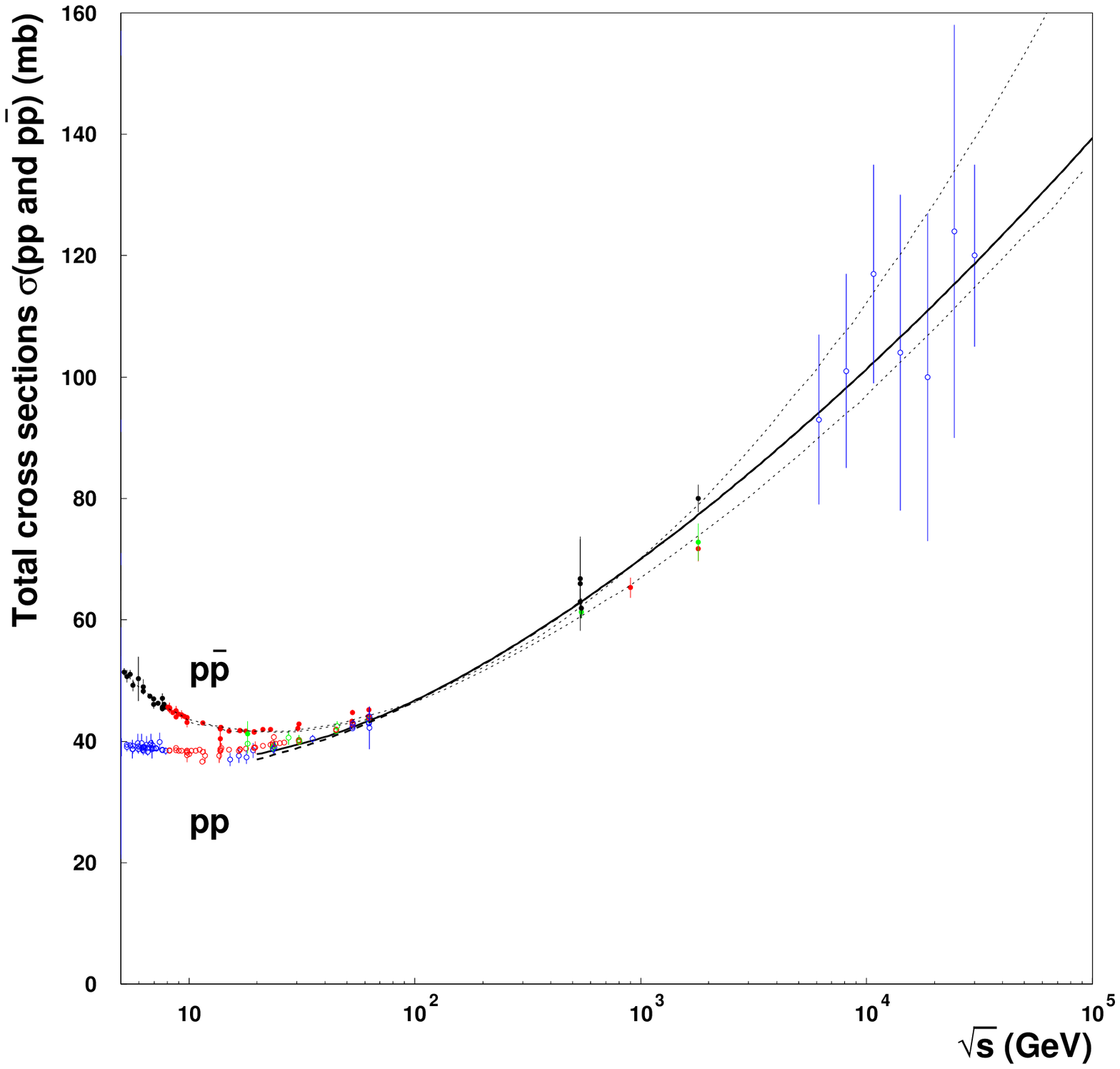}} 
\caption{ Our calculated $\sigma_{tot}$ for $\bar pp$ (solid line) and $pp$ (dashed line) 
are shown as functions of $\sqrt{s}$. Total cross sections for $\bar pp$  and $pp$ essentially overlap. 
The dotted lines represent the region of uncertainty for $\sigma_{tot}$
 obtained from dispersion relation calculations 
\cite{[11]}. Experimental data are given for comparison.
\label{Fig.1}}
\end{figure}

\begin{figure}[H]
\centering {\epsfxsize=170mm
\epsffile{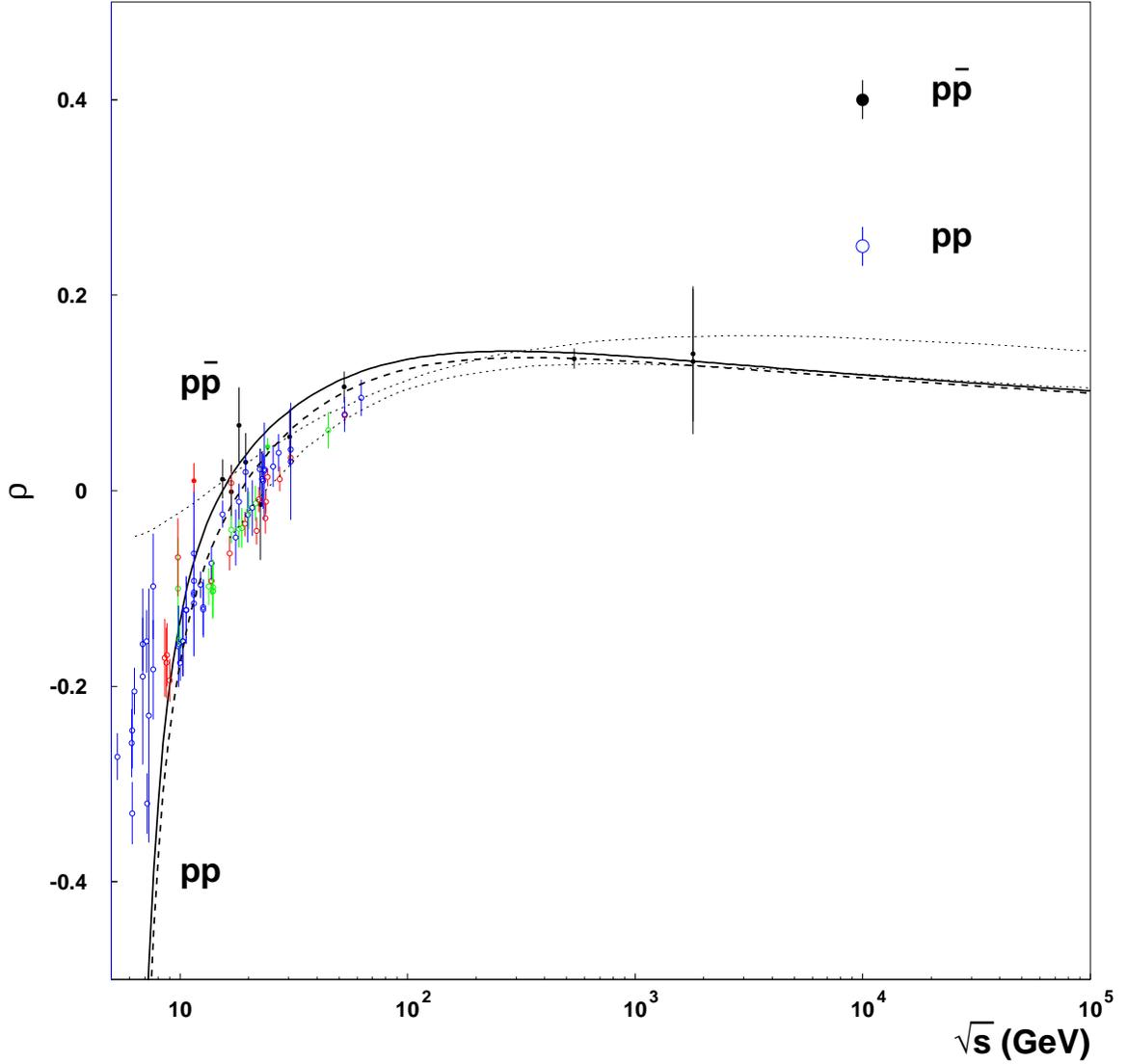}} 
\caption{ $\rho (s) = {\rm Re} T(s,0) / {\rm Im} T(s,0)$ calculated by us is shown as a function of 
$\sqrt{s}$. Solid curve corresponds to $\bar pp$ and dashed curve to $pp$. The dotted lines 
represent the region of uncertainty for $\rho (s)$ obtained from dispersion relation calculations 
\cite{[11]}. Experimental data are given for comparison.
\label{Fig.2}}
\end{figure}

\begin{figure}[H]
\centering {\epsfxsize=170mm
\epsffile{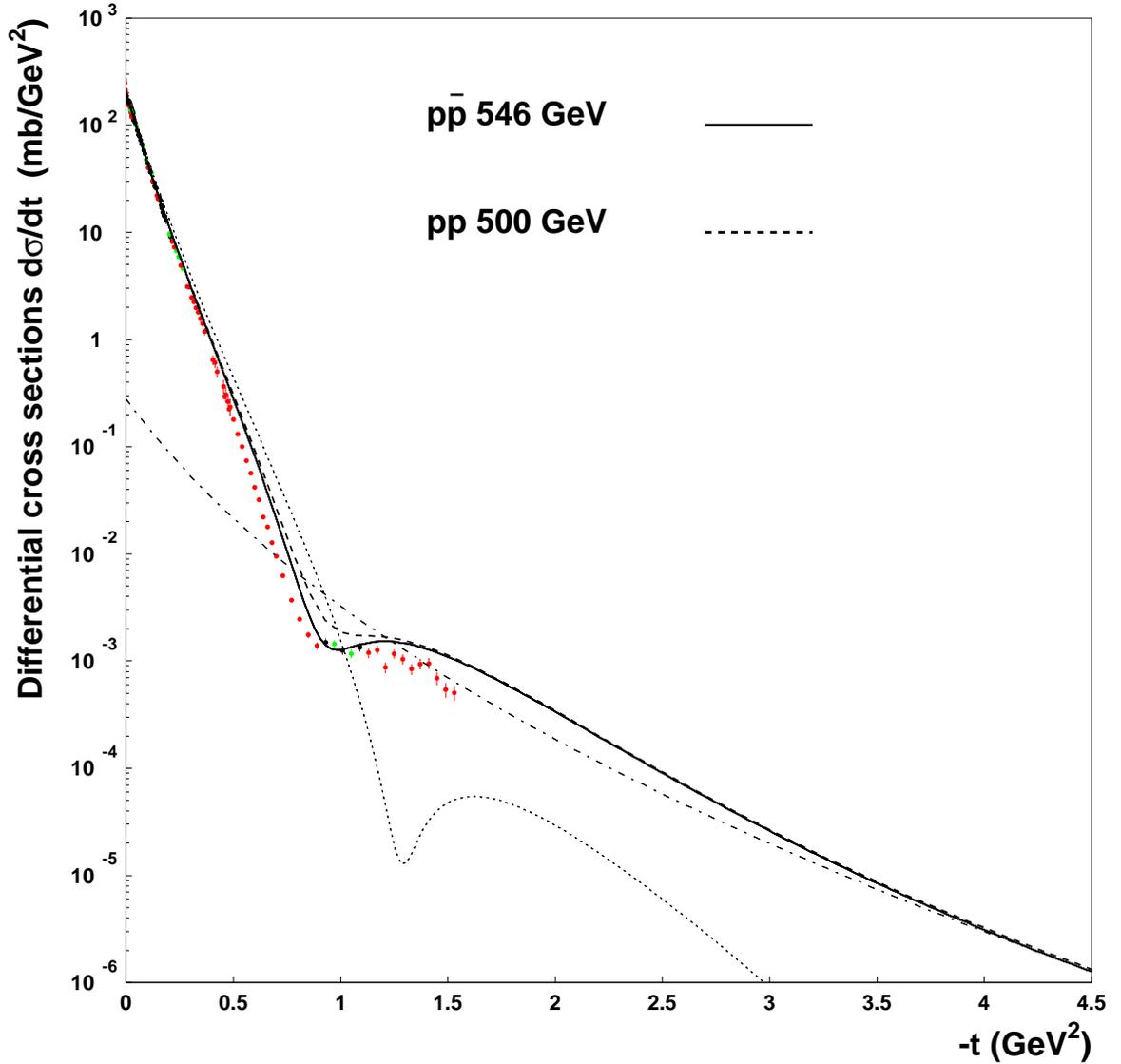}} 
\caption{Solid curve represents our calculated $d\sigma/dt$ for $\bar pp$ at 
 $\sqrt{s}= 546\;{\rm GeV}$. Experimental data 
are from Bozzo et al. \cite{[2]}. Dotted curve represents $d\sigma/dt$ due to diffraction alone, 
while dot-dashed curve represents $d\sigma/dt$ due to hard-scattering alone. 
The thick-dashed curve shows our predicted $d\sigma/dt$ for $pp$ elastic scattering at 
$\sqrt{s}= 500\;{\rm GeV}$, which will be measured at RHIC in the small $|t|$ region \cite{[12]}.
\label{Fig.3}}
\end{figure}

     In Fig.\ref{Fig.1}, our calculated $\sigma_{tot}(s)$ is shown as a function of $\sqrt{s}$ (solid curve). We find 
$\sigma_{tot}(s)$ for $\bar pp$  and $pp$ essentially overlap. 
The dotted lines represent the region of uncertainty for $\sigma_{tot}(s)$ obtained by Augier et al. 
\cite{[11]} from dispersion relation calculations \cite{[16]}. In Fig.\ref{Fig.2}, we show our calculated 
$\rho (s)$ as 
a function of $\sqrt{s}$ (solid curve: $\bar pp$ , dashed curve: $pp$ ). As in Fig.\ref{Fig.1}, the
 dotted lines 
represent the region of uncertainty for $\rho(s)$ obtained by Augier et al. \cite{[11]}. Experimental 
data are shown for comparison with the theoretical calculations. In Fig.\ref{Fig.3}, we show our calculated 
$d\sigma/dt$ for $\bar pp$ at $\sqrt{s}=546\;{\rm GeV}$ (solid curve) together 
with the experimental data of 
Bozzo et al. \cite{[2]}. Also shown are $d\sigma/dt$ due to diffraction alone 
(dotted curve) and due to hard scattering alone (dot-dashed curve). As in earlier calculations, 
there is destructive interference between diffraction amplitude and the hard scattering amplitude 
resulting in a dip at $|t|\approx 1.0\; {\rm GeV^2}$. The thick-dashed curve in Fig.\ref{Fig.3} shows our 
predicted $d\sigma/dt$ for $pp$ elastic scattering at $\sqrt{s}=500\;{\rm GeV}$, which will 
be measured at RHIC in the small $|t|$ region \cite{[12]}.

\begin{figure}[H]
\centering {\epsfxsize=170mm
\epsffile{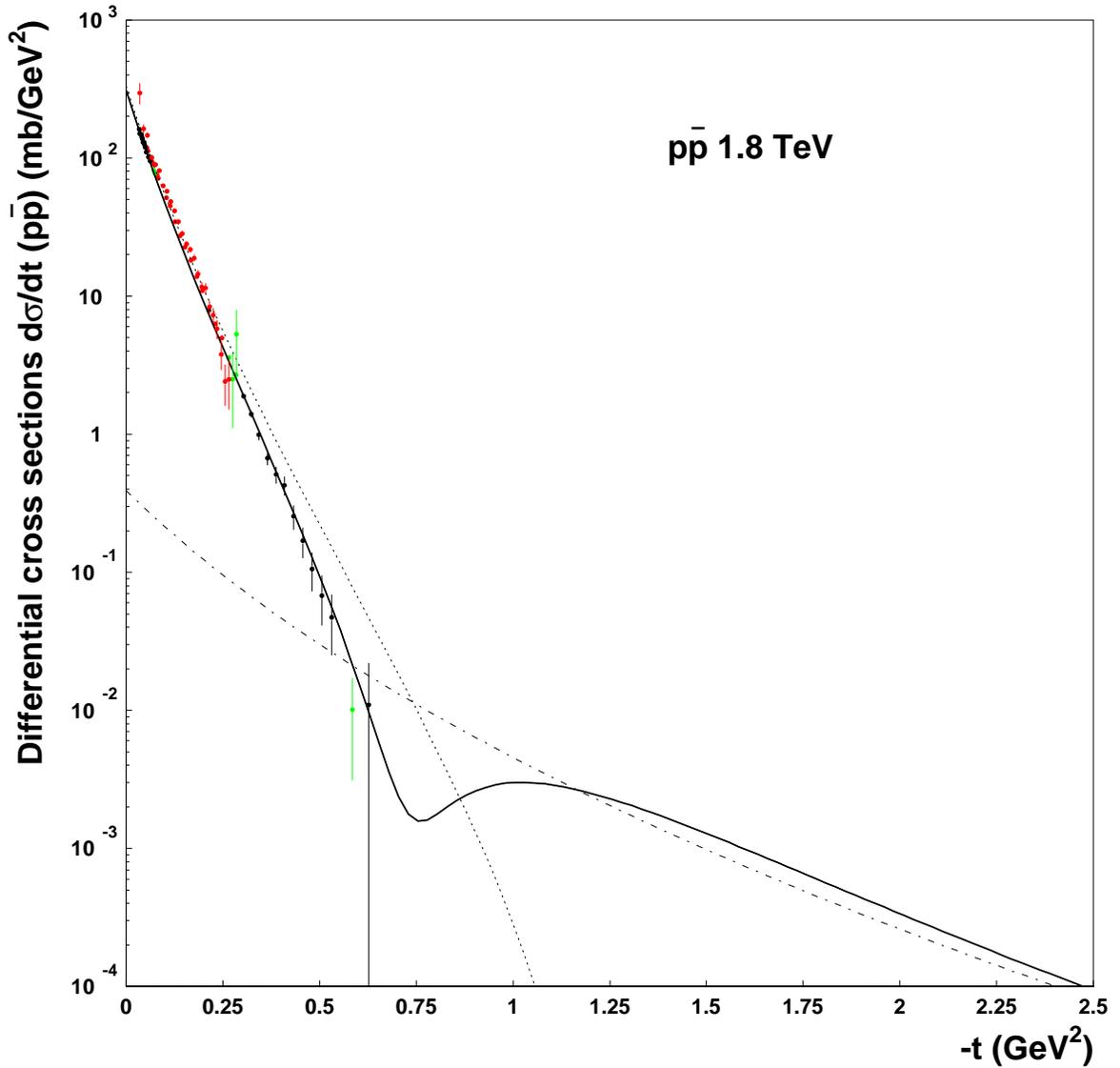}}
\caption{ Solid curve represents our predicted $d\sigma/dt$ for 
$\bar pp$ at $\sqrt{s}= 1.8\;{\rm TeV}$ with parameters 
determined from Figs.\ref{Fig.1},\ref{Fig.2}, and \ref{Fig.3}. 
Experimental data are 
from Amos et al. \cite{[3]} and Abe et al. \cite{[3a]}.
 Dotted curve represents $d\sigma/dt$ due 
to diffraction alone; dot-dashed curve represents $d\sigma/dt$ due
 to hard-scattering alone.
\label{Fig.4a}}
\end{figure}

\begin{figure}[H]
\centering {\epsfxsize=170mm
\epsffile{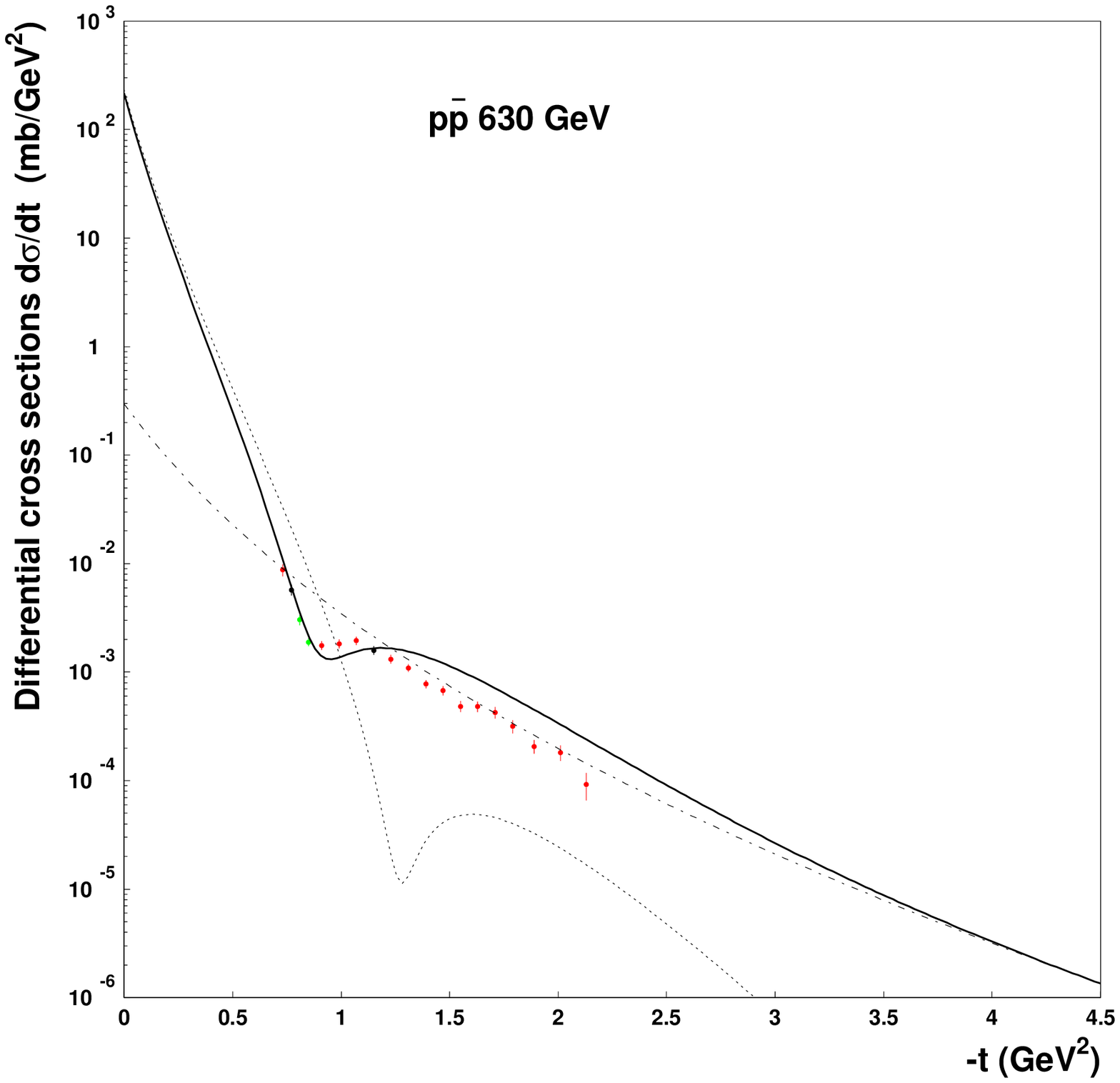}}
\caption{ 
As in Fig.\ref{Fig.4a}, solid curve represents our predicted $d\sigma/dt$ for
$\bar pp$
at $\sqrt{s}= 630$ GeV with parameters 
determined from Figs.\ref{Fig.1},\ref{Fig.2}, and \ref{Fig.3}. 
Experimental data are from Bernard et al. \cite{[15a]}. 
 Dotted curve represents $d\sigma/dt$ due 
to diffraction alone; dot-dashed curve represents $d\sigma/dt$ due
 to hard-scattering alone.
\label{Fig.4b}}
\end{figure}

With the parameters determined from $\sigma_{tot}(s)$ (Fig.\ref{Fig.1}),
$\rho (s)$ (Fig.\ref{Fig.2}), and $d\sigma/dt$ at 546 GeV (Fig.\ref{Fig.3}), 
we have asked ourselves: What are our predictions for $d\sigma/dt$ at $\sqrt{s}= 1.8$ TeV and $\sqrt{s}= 630$ GeV, 
and how do they compare with the existing experimental measurements? 
The results are shown in Figs.\ref{Fig.4a} and \ref{Fig.4b}. As can be seen, 
 our description of the differential cross section at 
1.8 TeV is quite satisfactory. 
As for the differential cross section at 630 GeV, we notice 
that the predicted $d\sigma/dt$ for $|t| > 1 {\rm GeV^2}$ is 
higher than the large $|t|$ data \cite{[15a]}. This reflects the fact 
that non-asymptotic terms are present in our analysis 
and contribute non-negligibly to  $d\sigma/dt$ for large $|t|$. 
This is also seen in Fig.\ref{Fig.3} for $\bar pp$ differential cross section 
at 546 GeV. Our model describes the asymptotic behavior and the approach 
to the asymptotic behavior of the elastic scattering amplitude. But, 
it cannot exclude the possibility of non-asymptotic terms contributing 
at some finite high energy such as 630 GeV and 546 GeV. The fact that 
our hard scattering contribution alone accounts for the large $|t|$
 differential cross section at 630 GeV and 546 GeV supports 
the point that the core-core scattering is showing up. Furthermore, 
it is because of the possibility of non-asymptotic terms being present 
that we  determine our parameters not by $\chi^2$ analysis, but by 
requiring that they describe well $\sigma_{tot}(s)$ , $\rho (s)$ and 
$d\sigma/dt$ at 546 GeV. Finally, as far as odderon  
contribution to $d\sigma/dt$ is concerned, we find it quite small and 
only noticeable in the dip region.

\begin{figure}[H]
\centering {\epsfxsize=170mm
\epsffile{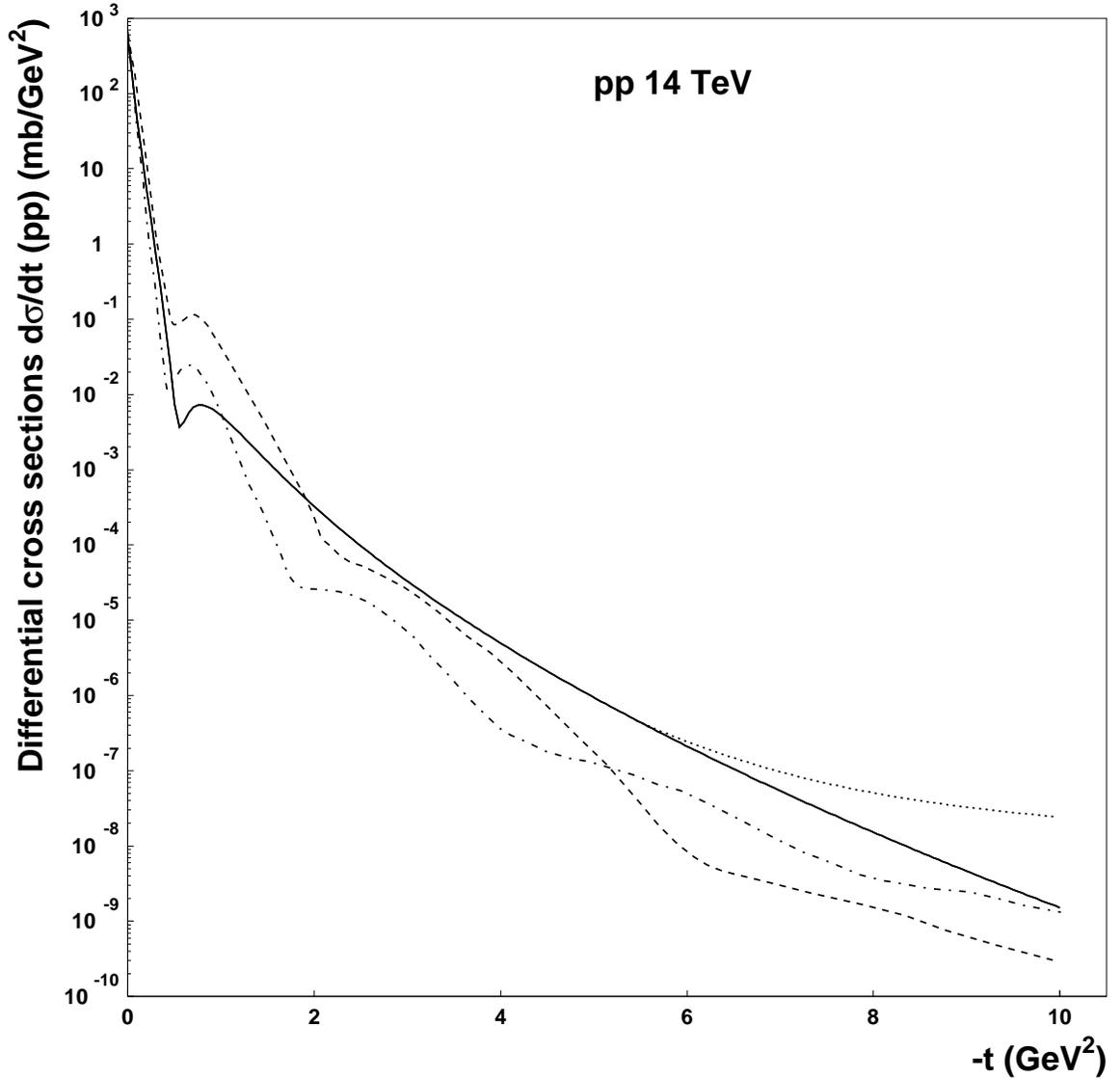}}
\caption{Solid curve shows our predicted $pp$ elastic differential cross section 
at LHC at the c.m.energy 14 TeV. Also shown for comparison are $d\sigma/dt$ 
predicted by the impact-picture model at $\sqrt{s}= 14\;{\rm TeV}$ (dashed curve) and by the Regge
 pole-cut model at $\sqrt{s}= 16\;{\rm TeV}$ (dot-dashed curve) \cite{[4]},
\cite{[5]}, \cite{[17]}. The dotted line represents schematically a change 
in the behavior of $d\sigma/dt$ predicted by our model, because of 
transition from the nonperturbative regime to the perturbative regime. 
\label{Fig.4}}
\end{figure}

    In Fig.\ref{Fig.4}, we show our predicted $pp$ elastic differential cross section at 
the LHC c.m. energy 14 TeV (solid curve). Also shown for comparison are $d\sigma/dt$ predicted by the 
impact picture model at $\sqrt{s}=14\;{\rm TeV}$ (dashed curve) and ${d\sigma}/{dt}$ predicted by the Regge 
pole-cut model at $\sqrt{s}=16\;{\rm TeV}$ (dot-dashed curve) \cite{[4]},\cite{[5]},\cite{[17]}. 
We find that the  real part of the diffraction amplitude has a zero at $|t| = 0.087\; {\rm GeV^2}$ 
as expected from Martin's theorem \cite{[14]}.  The dotted line in Fig.\ref{Fig.4} represents schematically 
the expected change in our model in the behavior of ${d\sigma}/{dt}$ from Orear fall-off: 
$d\sigma/dt\sim e^{-a\sqrt{|t|}}$ to a power fall-off: $d\sigma/dt\sim 1/t^{10}$ due to quark-quark
 scattering \cite{[8]},\cite{[18]}. It corresponds to a transition from the nonperturbative 
regime to the perturbative regime and should appear as a distinct change in the slope of 
$d\sigma/dt$ at large $|t|\sim 8\;{\rm GeV^2}$ \cite{[19]}. We further note that in our model ${d\sigma}/{dt}$ 
falls off smoothly for $|t|\gtrsim 1.5\;{\rm GeV^2}$. On the other hand, impact-picture model 
and Regge pole-cut model predict in this $|t|$ region oscillatory behavior typical of diffraction models 
\cite{[17]}.

    In conclusion, we have been able to extend previous $\bar pp$, $pp$ elastic $d\sigma/dt$ calculations
 at ISR and SPS Collider energies \cite{[7]} to the high energy asymptotic region. 
This has been possible by parameterizing the energy-dependent parameters suitably consistent with their crossing properties and by requiring that the model describes satisfactorily: 
(1) the high energy behavior of $\sigma_{tot}(s)$ and $\rho(s)$ as given by dispersion relation 
calculations, and 
(2) the experimentally measured elastic $\bar pp$ differential cross section at $\sqrt{s}=546\;{\rm GeV}$. 
We then 
obtain a quantitative prediction of $pp$ elastic  $d\sigma/dt$ at LHC at the c.m. energy 14 TeV. 
Our phenomenological investigation of elastic $\bar pp$ and $pp$   scattering is based on 
nucleon having an outer cloud of quark-antiquark condensed state, an inner core of topological 
baryonic charge, and a much smaller quark-bag of radius $\lesssim0.1$ F \cite{[8]}. Correspondingly, in $d\sigma/dt$
 we have diffraction  scattering dominating the small $|t|$ region, 
hard scattering dominating the intermediate $|t|$ region ($1.5\lesssim  |t| \lesssim 8.0\;{\rm GeV^2}$) 
and quark-quark scattering taking over in the large $|t|$ region 
(see Fig.\ref{Fig.4}). The $|t|$ dependence 
of these three regions are very different, so that accurate measurement of $d\sigma/dt$ in the TOTEM experiment 
\cite{[9]} will be able to identify these three regions. We also note that the underlying 
field theory model of the above nucleon structure based on gauged linear $\sigma$-model indicates 
that the transition from the nonperturbative regime of diffraction and hard scattering 
to the perturbative regime of quark-quark scattering is a chiral phase transition \cite{[8]}. 
Therefore, observation of a distinct change in the slope of $d\sigma/dt$  at large 
$|t| \gtrsim 8.0\;{\rm GeV^2}$ will provide an important signature of this phase transition.

The authors would like to thank Enrico Predazzi for his interest and comments. RJL would like 
to thank Ben Luddy for his programming assistance. MMI would like to thank Mike Albrow for bringing Fermilab E-811 Collaboration result to his attention.


\newpage

\end{document}